\documentclass[manuscript]{aastex}

\newcommand{\mb}{\mathbf}
\slugcomment{to be submitted}

\shorttitle{MDR-Based Approach to Coronal Magnetic Field
Extrapolation} \shortauthors{Hu et al.}

\begin{document}


\title{A Practical  Approach to Coronal Magnetic Field Extrapolation Based on the Principle of Minimum Dissipation Rate}


\author{Qiang Hu and B. Dasgupta}
\affil{Institute of Geophysics and Planetary Physics, University
of California,
    Riverside, CA 92521}
\email{qiang.hu@ucr.edu}

\author{D. P. Choudhary}
\affil{Department of Physics and Astronomy, California State
University, Northridge, CA 91330 }

\and

\author{J. B\"{u}chner}
\affil{Max-Planck-Institut f\"ur Sonnensystemforschung,
Max-Planck-Str. 2, 37191, Katlenburg-Lindau, Germany }


%\altaffiltext{1}{Visiting Astronomer, Cerro Tololo Inter-American Observatory.
%CTIO is operated by AURA, Inc.\ under contract to the National Science
%\altaffiltext{2}{Society of Fellows, Harvard University.}
%\altaffiltext{3}{present address: Center for Astrophysics,
%    60 Garden Street, Cambridge, MA 02138}
%\altaffiltext{4}{Visiting Programmer, Space Telescope Science Institute}
%\altaffiltext{5}{Patron, Alonso's Bar and Grill}


\begin{abstract}
We present a newly developed approach to solar coronal magnetic
field extrapolation from vector magnetograms, based on the
Principle of Minimum Dissipation Rate (MDR). The MDR system was
derived from a variational problem that is more suitable for an
open and externally driven system, like the solar corona. The
resulting magnetic field equation is more general than force-free.
Its solution can be expressed as the superposition of two linear
(constant-$\alpha$) force-free fields (LFFFs) with distinct
$\alpha$ parameters, and one potential field. Thus the original
extrapolation problem is decomposed into three LFFF
extrapolations, utilizing boundary data. The full MDR-based
approach requires  two layers of vector magnetograph measurements
on solar surface, while a slightly modified practical approach
only requires one. We test both approaches against 3D MHD
simulation data in a finite volume. Both yield  quantitatively
good results. The errors in the magnetic energy estimate are
within a few percents. In particular, the main features of
relatively strong perpendicular current density structures,
representative of the non-force freeness of the solution, are well
recovered.
\end{abstract}



\section{Introduction}
Plasma relaxation processes are ubiquitous in astrophysical and
laboratory plasmas. One well-known approach is to invoke the
variational principle, minimizing the total magnetic field energy
subject to the constraint of constant total magnetic helicity
\citep{taylor74}. Such an approach yields the magnetic field
equation for a relaxed plasma state, in which the Lorentz force
vanishes, i.e., the so-called linear force-free field (LFFF) with
a constant $\alpha$ parameter,
\begin{equation}
\nabla\times\mb{B}=\alpha\mb{B}.
\end{equation}\label{lfff}
Although the above equation~(\ref{lfff}) has been widely applied
in modeling magnetic field configurations in laboratory plasma
devices and in solar atmosphere, it is found inadequate in certain
situations \citep[e.g.,][]{gary01, met95}. In particular,
\citet{ama00}, among others, showed by 3D numerical MHD simulation
that in certain solar physics situation, after initial helicity
drive, the final ``relaxed state is far from the constant-$\alpha$
linear force-free field that would be predicted by Taylor's
conjecture" with helicity conserved. They suggested to derive
alternative variational problem.

An alternative approach, also based on the variational principle,
was recently developed for an open system with external drive, and
applied to theoretical investigation of the solar arcade
structures with flow \citep{BJ04, Bha07}. It was later applied to
develop a new approach to the extrapolation of solar coronal
magnetic field in non-force free state \citep{hu06, hu06sol,
hu07aip}. The core of the approach is to construct a variational
problem based on the Principle of Minimum Dissipation Rate (MDR),
which states ``The steady state of an irreversible process is
characterized by a minimum value of the rate of entropy
production" \citep{pri47}. In most cases, the entropy production
rate is equivalent to the energy dissipation rate. The basis is
the set of generalized momentum balance equations, and the system,
with flow, is always in broadly-defined force-balanced dynamic
equilibrium, satisfying the MHD equations \citep{BJ04}.  Prior
works based on MDR \citep{mon88, das98, das02} had successfully
shown that it was able to yield a pressure-balanced configuration,
supporting a finite plasma pressure gradient found in plasma
confinement devices. Recently a proof of MDR was rigorously sought
by 3D numerical simulations \citep{shk07}. We refer the readers to
the above referenced literatures for detailed description and
justification of the theoretical basis of MDR.

In short, analogous to the principle of minimum energy, but for a
more complex, open system like the solar corona, the MDR yields
the following equations for the magnetic field $\mb{B}$ and flow
vorticity $\vec\omega=\nabla\times\mb{v}$, $\mb{v}$ being the
plasma flow velocity \citep{BJ04, Bha07, hu06sol}:
\begin{eqnarray}
 \nabla\times\nabla\times {\bf B}+  a_1 \nabla\times {\bf B}+ b_1{\bf
 B}&=&\nabla \psi,  \label{nffmag} \\
\nabla\times\nabla\times {\vec \omega}+ a_2 \nabla\times {\vec
\omega} + b_2{\vec \omega}&=&\nabla \chi.    \label{nfflow}
\end{eqnarray} Here,  coefficients $a$ and $b$ are constants, and involve
the parameters of the system. The right-hand sides of both
equations are arbitrary, undetermined functions.

In what follows, we first briefly provide a heuristic
justification of the MDR in Section~\ref{just}, based on a simple
but standard analysis. We then present a brief but comprehensive
description of the full MDR-based coronal magnetic field
extrapolation approach in Section~\ref{fullapp}, including a new
test case study utilizing 3D MHD simulation data of a bright point
region, highlighting the non-force free features of the solution.
In Section~\ref{praapp}, we develop a practical approach that
requires only one single-layer vector magnetogram and show that it
achieves the same satisfactory results as the full approach for
the same test case. In the last Section~\ref{conclu}, we conclude
and discuss the significance and limitations of our approach.

\section{Heuristic Justification of the MDR}\label{just}
In a generally resistive plasma,  the {Lundquist number, $S$, is
defined as the ratio of the resistive time scale over the
Alfv\'{e}n time scale, $ S= \frac{\tau_R}{\tau_A}$
\citep[e.g.,][]{ort93}. The Lundquist number scales with
$L_0/\eta_0$, the ratio of the characteristic length over the
resistivity of the system. Therefore the number $S$ can be very
large when $L_0$ is large and/or plasma is highly conductive
($\eta_0$ is small). For example, it can reach $\sim 10^6 - 10^8$
in a hot fusion plasma, while for  solar coronal structures it
often exceeds $10^{12}$.

From standard procedures \citep[e.g.,][]{ort93}, the decay rates
of global magnetic helicity, $K$, and magnetic energy, $W$, are
given as follows (with the magnetic field being Fourier decomposed
as $\mb{B}=\sum_\mb{k} \mb{b}_\mb{k} \exp (i\mb{k}\cdot \mb{r}$)):
% and
%current density $\mathbf{j}$):
\begin{eqnarray}\label{hdecay1}
\frac{dK}{dt} &=&-\frac{2\eta}{S}\int_V {\bf j}\cdot{\bf B}dV\sim -\frac{2\eta}{S}\sum_k kb_k^2, \nonumber\\
R=\frac{d W}{dt}&=& -\frac{\eta}{S}\int_V j^2 dV\sim
-\frac{\eta}{S}\sum_k k^2 b_k^2. \nonumber
\end{eqnarray}
Taylor envisioned the relaxation process occurring as a result of
small scale turbulence, with $S\gg 1$. Following \citet{ort93}, at
scale lengths for which the Fourier decomposition wavenumber $k$
is $\sim S^{1/2}$,  the energy decay $dW/dt$ is $\sim O(1)$. But
for such scale length, the helicity decay rate $dK/dt$ is $ \sim
O(S^{-1/2})\ll1$. Thus, from this heuristic argument, we may
expect that small scale turbulence dissipate energy at a greater
rate than helicity.  In the same way, we can show that
$dR/dt$$\sim -\frac{2\eta^2}{S^2}\sum_k k^4 b_k^2$ goes at a much
faster rate that $dK/dt$. Furthermore, a recent 3D time-dependent
simulation of MHD fluids proves this argument \citep{shk07}. It is
shown that the energy dissipation rate, $\int \eta j^2 dV$, decays
the fastest towards a minimum state. We again refer readers to
\citet{shk07} for a complete description of the 3D numerical
simulation results.

\section{The Full MDR-Based Approach to Coronal Magnetic Field
Extrapolation}\label{fullapp} The full MDR-based approach to
coronal magnetic field extrapolation has been developed and tested
against analytic models in \citet{hu06sol, hu07aip}, based on the
general equation yielded from MDR. The following equation results
by taking an extra curl on both sides of Eq.~(\ref{nffmag}):
 \begin{equation}
\nabla\times \nabla\times\nabla\times {\bf B}+  a_1
\nabla\times\nabla\times {\bf B}+ b_1\nabla\times{\bf
 B}=0.  \label{nff3}
\end{equation}
The aim is to solve the above equation in a finite volume
utilizing boundary data. The key is that one exact solution to
Eq.~(\ref{nff3}) exists and can be expressed as the superposition
of three LFFFs. Each satisfies Eq.~(\ref{lfff}) with distinct
$\alpha$ parameters. For the sake of completeness, we briefly
reiterate the procedures of the full approach in the following
subsection.

\subsection{Procedures} \label{proc}
The exact solution to Eq.~(\ref{nff3}) is written
\begin{equation}
\mb{B}=\mb{B}_1+\mb{B}_2+\mb{B}_3, \label{B3}
\end{equation}
with $\nabla\times \mb{B}_i=\alpha_i \mb{B}_i$, $i=1,2,3.$
Subsequently, one obtains (taking curls on both sides of
Eq.~(\ref{B3}))
\begin{displaymath}
\left( \begin{array}{c} \mb{B} \\ \nabla\times\mb{B} \\
\nabla\times\nabla\times\mb{B}\end{array} \right)
=\mathcal{V}\left( \begin{array}{c} \mb{B}_1 \\ \mb{B}_2 \\
\mb{B}_3 \end{array} \right),
\end{displaymath}
where the matrix $\mathcal{V}$ is a (transposed) Vandermonde
matrix composed of constant elements $\alpha_i^{j-1}, i,j=1,2,3$.
It is guaranteed inversible provided that the parameters
$\alpha_i$, i=1,2,3, are distinct. Therefore, it follows
\begin{equation}
%\begin{displaymath}
\left( \begin{array}{c} \mb{B}_1 \\ \mb{B}_2 \\
\mb{B}_3 \end{array} \right)=\mathcal{V}^{-1}\left( \begin{array}{c} \mb{B} \\ \nabla\times\mb{B} \\
\nabla\times\nabla\times\mb{B}\end{array} \right) .\label{bvc}
%\end{displaymath}
\end{equation}
The above equations provide the bottom boundary condition for each
LFFF $\mb{B}_i$, given vector magnetograph measurements $\mb{B}$
at certain bottom levels on the solar surface. In order to
evaluate the double-curl term at the bottom boundary, ideally at
least two levels of vector magnetograms have to be utilized and
only the normal component of $\mb{B}_i$ perpendicular to the
bottom boundary can be obtained \citep{hu06sol}. Then each LFFF
$\mb{B}_i$ is extrapolated into a finite volume above the bottom
boundary using the FFT-based method by \citet{ali81} with known
$\alpha_i$ parameters for a Cartesian geometry. The summation of
the three LFFFs yields the final solution according to
Eq.~(\ref{B3}), a non-force free field in general.

It turns out that one of the $\alpha_i$ has to be zero and the
corresponding $\mb{B}_i$ becomes potential. The reason is simply
that the solution (\ref{B3}) has to satisfy Eq.~(\ref{nff3})
\citep[see][ and Section~\ref{praapp}]{hu06sol,hu07aip}. We
arbitrarily choose $\alpha_2=0$, then $a_1=-(\alpha_1+\alpha_3)$
and $b_1=\alpha_3\alpha_1$.

The determination of $\alpha_1$ and $\alpha_3$ is by
trial-and-error: enumerate all possible pairs of $\alpha_1$ and
$\alpha_3$, preferably within the limit of $\alpha_{max}$
\citep{gary89}, for each pair and $\alpha_2\equiv 0$, solve for
the transverse field $\mb{b}_{it}$ at the bottom boundary only,
then evaluate the deviation between the ``exact" (measured)
$\mb{B}_t$ and $\mb{b}_t=\sum_{i=1}^{3}\mb{b}_{it}$ by
\citep[e.g.,][]{sch06} \begin{equation}
E_n=\sum_{i=1}^M|\mb{B}_{t,i}-\mb{b}_{t,i}|/\sum_{i=1}^M|\mb{B}_{t,i}|,
\label{En} \end{equation} where $M=N^2$, the total number of grids
on the transverse plane, and the optimal pair
$(\alpha_1,\alpha_3)_{opt}$ is chosen such that the corresponding
$E_n^{opt}$ is minimum.

To summarize, the procedures of the full MDR-based approach to
coronal magnetic field extrapolation constitute the following two
steps:
\begin{enumerate}
\item[1.] Set $\alpha_2=0$ and search through LFFF parameters, $\alpha_1$ and
$\alpha_3$ ($\alpha_1 > \alpha_3$, due to the exchangeability of
subscripts, 1 and 3). Apply an LFFF solver to calculate the
transverse components ${\mathbf{b}}_{it}$, utilizing the normal
component $B_{in}$ (given by equations~(\ref{bvc})), $i=1,2,3$, at
bottom boundary only. A pair of optimal $\alpha_i,i=1,3$, is
found, for which the $E_n$, given by Eq.~(\ref{En}), is minimum. A
plot of $E_n$ distribution over the $(\alpha_1,\alpha_3)$
parameter space is prepared to show the goodness and uniqueness of
the solution.
\item[2.] Solve for $\mathbf{B}_1$ and $\mathbf{B}_3$ in a finite volume above the bottom boundary,
 for the optimal $\alpha_1$ and $\alpha_3$ found in Step~1,
 respectively, and the potential field $\mb{B}_2$ as well. Obtain
$\mathbf{B}=\mathbf{B}_1+\mathbf{B}_2+\mb{B}_3$ (Eq.~(\ref{B3})).
%If
%certain plasma information is available, evaluate plasma
%pressure/density/temperature profile based on appropriate MHD
%equations.
\end{enumerate}

The run time of the approach based on the FFT algorithm solving
each LFFF is approximately in the order of $P\cdot N^2\log_2 N$,
largely dependent on the number of search grids, $P$, on
$(\alpha_1,\alpha_3)$ in Step 1. However, this process can be
easily parallelized by simply dividing the search domain into
individual pieces and assigning each piece to a separate
processor. One can also start with a coarse grid, then allocate a
small but finer grid around a minimum, thus the total number of
iterations can be effectively reduced.

\subsection{A Test Case Study Using Numerical Simulation
Data}\label{test} Several analytic solutions have been utilized to
test the full approach outlined in the previous subsection and the
results showed the approach was able to recover the solution in a
finite volume to certain degree of satisfactory accuracy for
several cases \citep{hu06sol, hu07aip}. Here in order to simulate
a case of real magnetograph measurements as closely as possible,
we employ 3D MHD simulation data surrounding a region of an X-ray
Bright Point (BP) \citep{bro01} from \citet{jbn05} \citep[see
also,][]{ott07,jb05,jb06}. The simulation started with actual
solar magnetic field measurements, and was continuously driven by
photospheric motion inferred from solar observations.

The data were provided in a finite Cartesian volume, $x\times y
\times z$=$128\times 128\times 63$, with $z$ being the vertical
dimension along the normal direction, at one chosen moment during
the 3D dynamic MHD simulation. The bottom two layers of vector
magnetic field data are utilized to generate the bottom boundary
($z=0$) conditions by Eqs.~(\ref{bvc}). The two steps are carried
out as described in Section~\ref{proc}. The distribution of $E_n$
in $(\alpha_1,\alpha_3)$ parameter space is shown in
Figure~\ref{fig1}. The minimum value is $E_n^{opt}\approx 0.30$,
with corresponding $(\alpha_1,\alpha_3)_{opt}\approx$(0.00156,
0.000779) (dimensionless), as denoted by the + sign in
Figure~\ref{fig1}. Fairly significant amount of uncertainty
associated with this optimal pair exists. However, further
analysis shows that choosing any other pair of
$(\alpha_1,\alpha_3)$ within the innermost contour does not
significantly change the extrapolation results. Such similar
pattern and small $\alpha$ values were also observed in
\citet{hu06sol}.

The corresponding optimal transverse magnetic field vectors
$\mb{b}_t$ at $z=0$, are shown in Figure~\ref{fig2}, together with
the ``exact" solution $\mb{B}_t$ from the simulation data and the
corresponding potential field results using the normal component
$B_z$ only \citep{ven89}. Our solution agrees very well with the
``exact" one, whereas the potential field result exhibits apparent
deviations.

The final extrapolation results are shown in Figure~\ref{fig3}, in
the form of 3D field line plots. The bottom image shows the normal
component distribution at $z=0$, with two major polarities in the
center corresponding to the location of BP. The field lines from
our result demonstrate a good deal of similarity (nearly identical
in the upper left part) to the ``exact" solution, indicating
good-quality recovery of the magnetic field in the finite volume
from the bottom two layers of data. This judgement is supported by
the set of quantitative measures, so-called figures of merit,
adopted from \citet{sch06}, given in Table~\ref{table1}. They are
used to evaluate, quantitatively, the agreement between the exact
vector field $\mb{B}$ and the corresponding extrapolated field
$\mb{b}$ in a volume (columns 2-5). The 2nd row in
Table~\ref{table1} lists the numbers for this calculation (result
1), and the last two rows list the values for a result identical
to the exact solution and a potential field extrapolation,
respectively. The last two columns show the energy estimates, the
magnetic energy ratio of the extrapolated field $\mb{b}$ over the
exact field $\mb{B}$, i.e., $\epsilon=\mb{b}^2/\mb{B}^2$, and the
energy ratio, $\epsilon_p$, that of $\mb{b}$ over the
corresponding potential field. For this case, several figures from
the potential field extrapolation are comparable with our results.
However, the energy estimates clearly mark the distinction: ours
only differs from the exact ones by $\sim -1\%$, whereas the
potential field results differ by as large as $\sim -16\%$.

To further characterize the non-force freeness of the solution,
which cannot be captured by either the potential or the force-free
extrapolation, and to better visualize the results, we calculate
the field-line integrated current density, $J=\int {j} d{l}$,
along individual field lines \citep[e.g.,][]{jb06}.
Figure~\ref{fig4} shows such quantity separated into field-aligned
($\parallel \mb{B}$) component, $J_\parallel$, and the component
perpendicular to $\mb{B}$, $J_\perp$, for the ``exact" solution. A
non-vanishing $J_\perp$ indicates the field is non-force free. Two
footpoints of one single field line, both rooted on the bottom
boundary, have the same value. Figure~\ref{fig5} shows the
corresponding $J_\parallel$ and $J_\perp$ for our result 1
discussed above. The distribution of $J_\parallel$ is well
recovered with concentrations at the two major polarities around
the center. The major features of $J_\perp$ at the same locations
are well recovered as well, but most of the remaining weak
structures are lacking, compared with the right panel in
Figure~\ref{fig4}. However, both $J_\perp$ and $J_\parallel$
enhancements near the lower boundary tracing back to one of the
strong polarities to the right, remain. It is indicative of strong
currents, implying the possible locations of current sheets that
are important for magnetic reconnection \citep{jb05,jb06}.

At present time, multiple ($\ge 2$) layers of vector magnetograms
are hardly available \citep[but see, e.g.,][]{met95}. Only the
photospheric vector magnetic fields are routinely observed.
Sometimes, the  chromospheric magnetic field line-of-sight (LOS)
components were inferred \citep[e.g.,][]{cho01}. When the
observations are near the disk center, the LOS component can be
used as the normal component $B_z$, then the term
$(\nabla\times\nabla\times\mb{B})_z$=$-\nabla^2 B_z$ may be
approximated by (denoting the distance between the chromosphere
and the photosphere by $\Delta z$),
\begin{equation}
\nabla^2 B_z\approx\nabla_t^2
B_z+2(B_z^{chrom}-B_z^{photo}-\frac{\partial B_z}{\partial
z}\Delta z)/\Delta z^2. \label{d2Bz}
\end{equation}
The first-order derivative $\partial B_z/\partial z$ is
approximated by the divergence free condition using transverse
magnetic field vectors at the photosphere. The 3rd row of
Table~\ref{table1} (result 2) lists the figures of merit of the
extrapolation result using the bottom boundary  magnetic field
vectors and only the $B_z$ component at one level immediately
above to approximate $\nabla^2 B_z$ by Eq.~(\ref{d2Bz}). It shows
somewhat deteriorated results, and the error in energy estimate is
$\sim +12\%$.

\section{A Practical Approach Utilizing Single-Layer Vector Magnetogram}\label{praapp}
As mentioned earlier, the multi-layer vector magentograms are not
routinely available. On the other hand, the high-quality
photospheric vector magnetograms become increasingly available. In
order to apply our method to actual measurements currently
available, we adapt our approach to employ only one single layer
vector magnetogram, at the expense of less-general, expected
limited applications, as to be described below.

From Eq.~(\ref{nffmag}), considering the arbitrary nature of
$\nabla\psi$, one may write one exact solution to (\ref{nffmag}),
\begin{equation}
\mb{B}=\mb{B}_1+\mb{B}_3+c\mb{B}_{pot},\label{B3c}
\end{equation}
where a constant, $c$, and the potential field, $\mb{B}_{pot}$,
obtained from the known normal field $B_z$ at the bottom boundary,
are introduced. Then it follows $\nabla\psi=b_1 c\mb{B}_{pot}$.
This is equivalent to the full approach in Section~\ref{fullapp},
except that the potential field $\mb{B}_2$ is chosen a special
form, $c\mb{B}_{pot}$, and is largely known subject to a
undetermined constant.

With an additional unknown parameter, $c$, the procedures are
similar to those outlined in Section~\ref{proc}, but for a
reduced, 2nd order system. The following equations, in place of
(\ref{bvc}), now provide the bottom boundary conditions for each
LFFF from one single layer vector magnetogram:
\begin{eqnarray}
(\alpha_3-\alpha_1)B_{1z}&=&\alpha_3 B'_z-(\nabla\times\mb{B})_z,
\nonumber \\
(\alpha_1-\alpha_3)B_{3z}&=&\alpha_1 B'_z-(\nabla\times\mb{B})_z,
\nonumber
\end{eqnarray}
with $B'_z=B_z-cB_{pot,z}$. In addition to the
$(\alpha_1,\alpha_3)$ search domain in Step~1, another dimension
in $c$ is added. A desirable choice of allowable $c$ values is
$c\in [-1,1]$, to limit the dominance of the potential field. The
approach first reported in \citet{hu06} represents the special
case, $c\equiv 0$.

We again test this practical approach against the same data set
utilized in Section~\ref{fullapp}, but only use the vector
magnetic field data on the bottom boundary, $z=0$. Only a limited
number of $c$ values are chosen, $c\in [-1,-0.5,0,0.5,1]$.  The
$E_n$ distribution at $c_{min}=-1.0$, which yields the minimum,
$E_n^{opt}\approx 0.22$, is shown in Figure~\ref{fig6}. Compared
with Figure~\ref{fig1}, the uniqueness of
$(\alpha_1,\alpha_3)_{opt}$ is much improved. The corresponding
$\mb{b}_t$ at $z=0$, and the 3D field line plot are shown in
Figures~\ref{fig7} and \ref{fig8}, respectively. The corresponding
figures of merit are given in the 4th row (result 3) of
Table~\ref{table1}. Good agreement with the exact solution is
achieved. The energy estimate has a $\sim -5\%$ error. The
field-line-integrated current densities shown in Figure~\ref{fig9}
are essentially the same as the previous ones obtained by the full
approach. All the main features discussed in Section~\ref{test}
are retained.

\section{Conclusions and Discussion}\label{conclu}
In conclusions, we develop an approach to extrapolate the coronal
magnetic field from vector magnetograms based on the Principle of
Minimum Dissipation Rate (MDR). Analogous to, but yet opposed to
the principle of minimum energy, the MDR yields a generally
non-force free magnetic field through a different variational
approach. The full MDR-based approach requires two layers of
vector magnetograms, while a practical approach, representing a
class of special solutions to the MDR system, requires only one
layer, which is more amiable for practical applications to
currently available data. A test case study using numerical
simulation data shows that both  approaches recover the solution
to a good degree of accuracy, as measured by a set of quantitative
measures. The errors in energy estimate are both within a few
percents. Moreover, the non-force free features of the solution,
mainly the strong perpendicular current density concentrations,
are well retained in the extrapolation results as well.

We hereby provide an alternative approach that is fast, and easy
to implement, and allows one to extrapolate coronal magnetic field
in a more general non-force free state with manageable effort in a
solo work. For instance, all the calculations reported herein
(128$\times$128$\times$63 grid) were performed on a
single-processor 2.8GHz PC in IDL within a reasonable time frame.
It potentially can be made much faster. It is apparent that the
Eq.~(\ref{nff3}) includes solutions to LFFF
($\nabla\times\mb{B}=\alpha\mb{B}$), for
$-\alpha_1=\alpha_3=\alpha$. It is not explicit whether this is
also the case for the nonlinear force-free field (NLFFF) when the
parameter $\alpha$ is allowed to vary, although several numerical
experiments indicated that this might be so \citep{hu06sol}
sometimes.

We are fully aware of the limitations of the method, due to the
fact that the currently employed LFFF solver originates from  an
ill-posed problem. It is adversely affected by the increasing
resolution of vector magetograms since $\alpha_{max}$ is inversely
proportional to $N$. This severely limits the search domain of
$\alpha_i$ that will yield unique physical solutions
\citep[e.g.,][]{ali81, gary89}. Recent progresses  in the
algorithm for NLFFF extrapolations \citep[e.g.,][]{son06}, in
addition to the existing LFFF algorithm \citep[e.g.,][]{abr96},
that modified the problem into a well-posed one, look promising
and may be adapted.

It is worth noting that our approach is wholly applicable to
extrapolating the plasma flow field, by simply replacing magnetic
field $\mb{B}$ with flow vorticity $\vec{\omega}$. However the
boundary conditions are probably harder to obtain since it is the
vorticity that is involved. A good start is to utilize the flow
field data contained in the same 3D MHD simulation test case. It
is not straightforward to obtain the plasma state
%by simply using
%$\nabla p=\mb{j}\times\mb{B}$,
since the flow field  has to be resolved as well, and to be
consistent, the full set of MHD equations has to be considered
\citep[see, e.g.,][]{mon88}. One may also foresee that extension
of this approach to the whole solar surface data is of great
significance. Further theoretical investigation is underway.

\acknowledgments

We are grateful to Drs. D. Shaikh, M.S. Janaki, and R.
Bhattacharyya for useful discussions. HQ and BD acknowledge NASA
LWS grant NNX07A073G for partial support.

%\appendix

\clearpage

\begin{figure}
\includegraphics{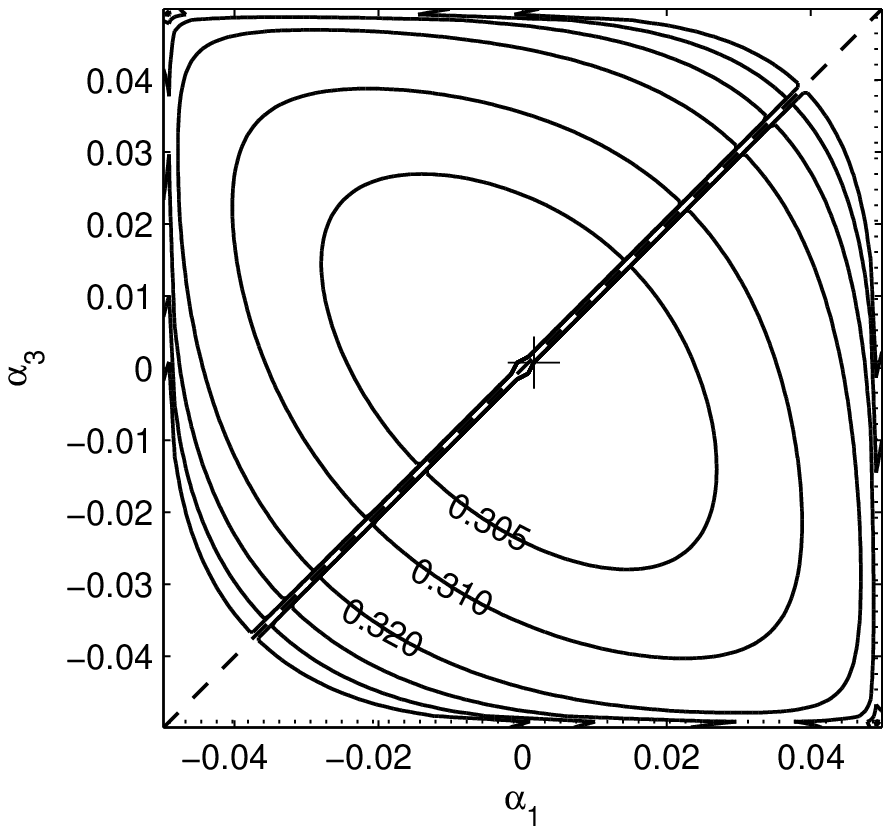} \caption{Contours of $E_n$ over the $(\alpha_1, \alpha_3)$ parameter space for the
 full MDR-based approach.
The unlabeled contour lines are of increment 0.01. The dotted
lines mark the limit where $|\alpha_{1,3}|=\alpha_{max}=2\pi/N$.
The plus sign denotes the location where $E_n$ is minimum. Due to
the interchangeability of $\alpha_1$ and $\alpha_3$, the plot is
symmetric about the dashed line, $\alpha_1=\alpha_3$.
}\label{fig1}
\end{figure}

%% Here we use \plottwo to present two versions of the same figure,
%% one in black and white for print the other in RGB color
%% for online presentation. Note that the caption indicates
%% that a color version of the figure will be available online.
%%

\begin{figure}
\includegraphics[width=\linewidth]{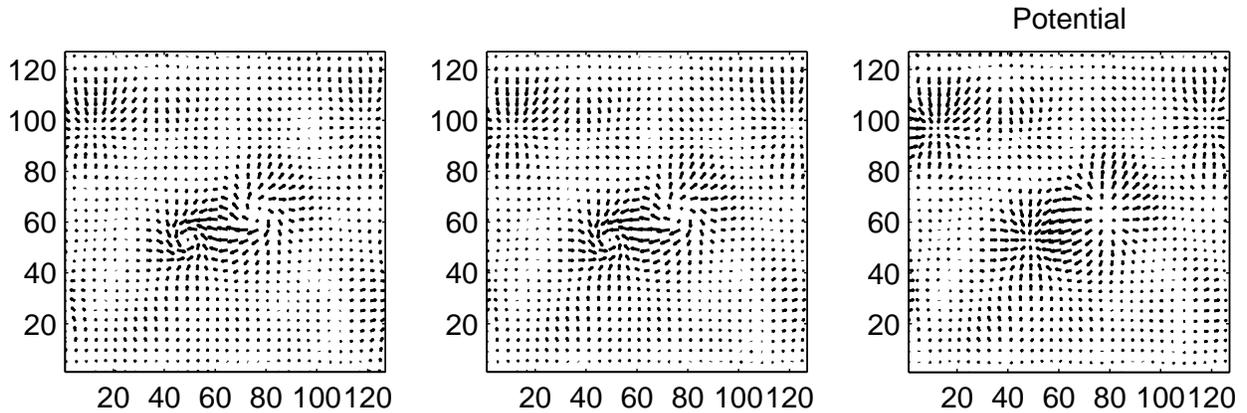} \caption{The transverse magnetic field vectors at $z=0$. {\em Left}: The exact distribution from
numerical simulation data. {\em Middle}: The derived optimal
result from the full MDR-based approach. {\em Right}: The result
from a corresponding potential field extrapolation.}\label{fig2}
\end{figure}

%% This figure uses \includegraphics to scale and rotate the still frame

\begin{figure}
\includegraphics{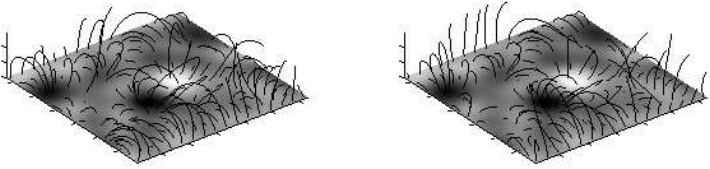}
\caption{3D field line plots of the exact solution (left panel)
and the corresponding extrapolated solution (right panel) from the
full MDR-based approach. All the field lines originate from the
same set of root points. The bottom image represents the normal
magnetic field component distribution on the bottom boundary, with
the gray scales ranging from strongly negative (black) to positive
(white). }\label{fig3}
\end{figure}
\begin{figure}
\includegraphics[]{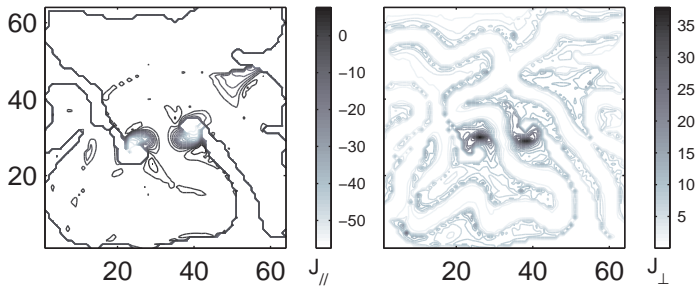}
\caption{Contour plots of field-line-integrated current densities
 from the exact solution. {\em Left}: The component
parallel to the magnetic field. {\em Right}: The component
perpendicular to the magnetic field. The gray colors show the
levels of the contour lines, as indicated by the
colorbars.}\label{fig4}
\end{figure}
\begin{figure}
\includegraphics[]{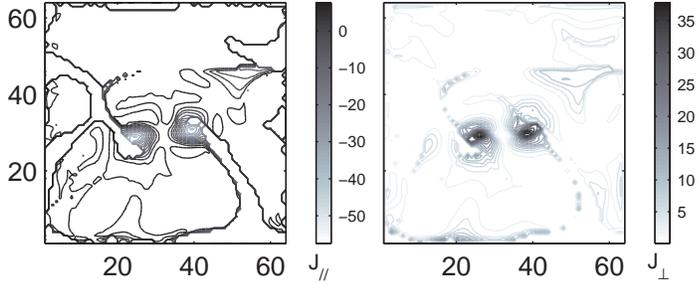}
\caption{The corresponding field-line-integrated current density
distributions obtained by the full MDR-based approach. Format is
the same as Figure~\ref{fig4}. So are the gray
scales.}\label{fig5}
\end{figure}
%\begin{figure}
%\includegraphics[]{Bt_JB128p.eps}
%\caption{}
%\end{figure}
\begin{figure}
\includegraphics{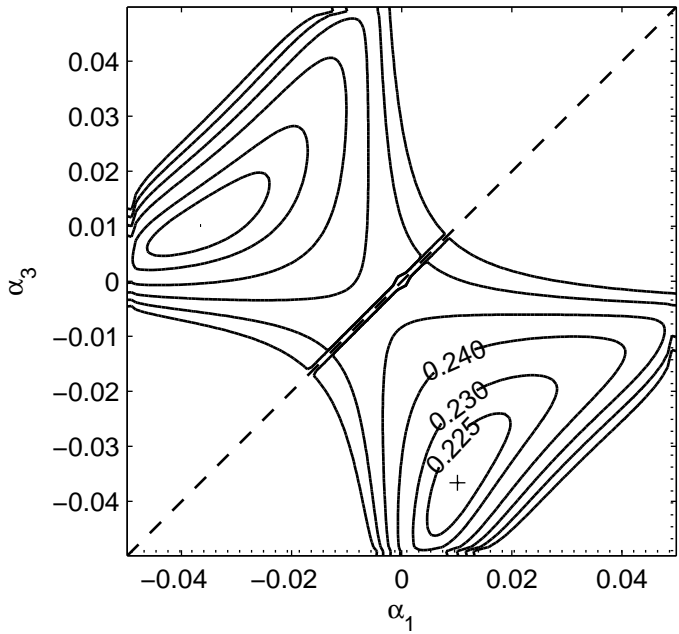}
\caption{The $E_n$ distribution for the practical MDR-based
approach. Format is the same as Figure~\ref{fig1}.}\label{fig6}
\end{figure}

\begin{figure}
\includegraphics[]{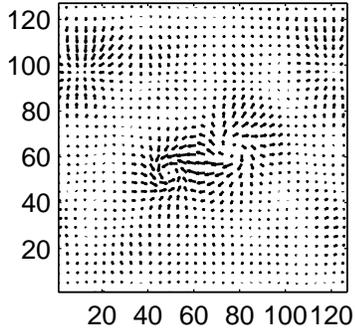}
\caption{The corresponding transverse magnetic field vectors at
$z=0$, obtained by the practical approach. }\label{fig7}
\end{figure}
\begin{figure}
\includegraphics[]{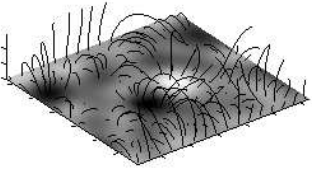}
\caption{The corresponding 3D field line plot, obtained by the
practical approach. Format is the same as
Figure~\ref{fig3}.}\label{fig8}
\end{figure}
\begin{figure}
\includegraphics[]{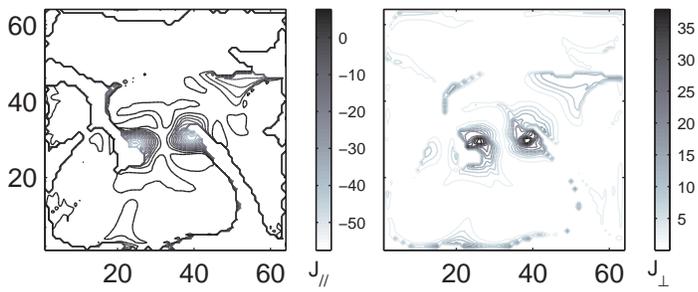}
\caption{The field-line-integrated current densities obtained by
the practical approach. Format is the same as Figure~\ref{fig4}.
}\label{fig9}
\end{figure}

%% If you are not including electonic art with your submission, you may
%% mark up your captions using the \figcaption command. See the
%% No more than seven \figcaption commands are allowed per page,
%% so if you have more than seven captions, insert a \clearpage

%% Tables should be submitted one per page, so put a \clearpage before

%% Two options are available to the author for producing tables:  the
%% deluxetable environment provided by the AASTeX package or the LaTeX
%% table environment.  Use of deluxetable is preferred.
%%

%% Three table samples follow, two marked up in the deluxetable environment,
%% one marked up as a LaTeX table.

%% In this first example, note that the \tabletypesize{}
%% command has been used to reduce the font size of the table.
%% We also use the \rotate command to rotate the table to
%% landscape orientation since it is very wide even at the
%% Note also that the \label command needs to be placed

%% This table also includes a table comment indicating that the full
%% version will be available in machine-readable format in the electronic

%% If you use the table environment, please indicate horizontal rules using
%% Do not put multiple tabular environments within a single table.
%% The optional \label should appear inside the \caption command.

\clearpage

\begin{table}
\begin{center}
\caption{Figures of merit (see \citet{sch06} and text for
definitions).\label{table1}}
\begin{tabular}{ccccccc}
\tableline\tableline Case BP & $C_{vec}$ & $C_{cs}$ & $E'_n$ &
$E'_m$ & $\epsilon$ & $\epsilon_p$ \\
\tableline
Our result 1\tablenotemark{a} &0.97&0.91 &0.74  &0.62 &\bf{0.99} &\bf{1.18}\\
Our result 2\tablenotemark{b} &0.91 &0.76 &0.53&0.27 &\bf{1.12}  &\bf{1.34}\\
Our result 3\tablenotemark{c} &0.97 &0.93  &0.76  &0.64 &\bf{0.95}  &\bf{1.14}\\
Exact  &1 &1  &1  &1 &\bf{1} & \bf{1.20}\\
Potential  & 0.95 &0.92 &0.70 &0.63 &\bf{0.84}  &\bf{1.00}\\
\tableline
\end{tabular}
%%% Any table notes must follow the \end{tabular} command.
\tablenotetext{a}{Results obtained by the full MDR approach (see
Section~\ref{fullapp}), where two bottom layers of vector magnetic
field data are utilized to derive the bottom boundary conditions.}
\tablenotetext{b}{Results obtained by the reduced bottom boundary
data (see Section~\ref{test}), where only the vector magnetic
field on the bottom layer, and the normal magnetic field component
on the layer immediately above are utilized.}
\tablenotetext{c}{Results obtained by the practical approach (see
Section~\ref{praapp}), where only the magnetic field vectors on
the bottom boundary are required.}
%\tablecomments{We can also attach a long-ish paragraph of
%explanatory material to a table.}
\end{center}
\end{table}


\begin{thebibliography}{}
\bibitem[Abramenko \& Yurchishin(1996)]{abr96} Abramenko, V.I., \&
Yurchishin, V. 1996, Sol. Phys., 168, 47
\bibitem[Alissandrakis(1981)]{ali81} Alissandrakis, C.E. 1981, {A\&A}, {{100}}, 197
\bibitem[Amari \&
Luciani(2000)]{ama00} Amari, T.,
    \& Luciani, J. F.  2000, \prl, 84, 1196
\bibitem[Bhattacharyya et al.(2007)]{Bha07}Bhattacharyya, R., Janaki, M.S., Dasgupta, B.,  \& Zank, G. 2007, {{Sol. Phys.}},   240, 63
%\bibitem{}Bhattacharyya, R.,  Janaki, M.S., and Dasgupta, B.: 2000, {\textit{Phys. Plasmas}}  {\bf 7},
%4801.
%\bibitem{} Bhattacharyya, R.,  Janaki, M.S., and
%Dasgupta, B.: 2003, {\textit{Plasma Phys. Control. Fusion}}
%{\bf{45}}, 63.
\bibitem[Bhattacharyya \& Janaki(2004)]{BJ04} Bhattacharyya, R., \& Janaki, M.S. 2004,  {{Phys. Plasmas}}, {{11}}, 5615
\bibitem[Brown et al.(2001)]{bro01} Brown, D.S. et al., 2001, Sol.
Phys., 201, 305
\bibitem[B\"uchner(2006)]{jb06} B\"uchner, J. 2006, \ssr, 122, 149
\bibitem[B\"uchner \& Nikutowski(2005)]{jbn05} B\"uchner, J., \& Nikutowski, B. 2005,
Procs. of the International Scientific Conference on Chromospheric
and Coronal Magnetic Fields, ESA SP-596
\bibitem[B\"uchner(2005)]{jb05} B\"uchner, J. 2005, Procs. of the
11th European Solar Physics Meeting, ESA SP-600
\bibitem[Choudhary et al.(2001)]{cho01} Choudhary, D.P., Sakurai, T., \& Venkatakrishnan, P. 2001,  \apj, {{560}}, 439
    \bibitem[Dasgupta et al.(1998)]{das98} Dasgupta, B.,  Dasgupta, P.,   Janaki, M.S.,  Watanabe, T.,
    \&
Sato, T. 1998, \prl,  {{81}},  3144
\bibitem[Dasgupta et al.(2002)]{das02} Dasgupta, B.,   Janaki, M.S., Bhattacharyya, R.,  Dasgupta, P.,
Watanabe, T.,  \&  Sato, T. 2002,  {\pre}, {{65}}, 046405.
%\bibitem{}  Gary, G. A.,(2001), Plasma beta above a solar active
%region: Rethinking the paradigm, Sol. Phys., {\bf{203}}, 71 .
%\bibitem{} Gary, G. A., and D. Alexender, (1999), Constructing the
%Coronal Magnetic Field By Correlating Parameterized Magnetic Field
%Lines With Observed Coronal Plasma Structures, Sol. Phys.,
%{\bf{186}}, 123.
\bibitem[Gary(1989)]{gary89} Gary, G.A. 1989,  {\apjs}, {{69}}, 323
\bibitem[Gary(2001)]{gary01} Gary, G. A.  2001, { Sol. Phys.},
    203, 71
    \bibitem[Hu \& Dasgupta(2006)]{hu06}Hu, Q., \& Dasgupta, B. 2006,  {\grl,} {{33}}, L15106
    \bibitem[Hu \& Dasgupta(2007)]{hu06sol}Hu, Q., \& Dasgupta, B. 2007,  {Sol. Phys.,}
    submitted
    \bibitem[Hu et al.(2007)]{hu07aip}Hu, Q., Dasgupta, B., \& Choudhary, D.P.  2007,
    AIP CP932, 376
\bibitem[Metcalf et al.(1995)]{met95} Metcalf, T. R. et al., 1995, \apj, 439,
474
\bibitem[Montgomery \& Phillips(1988)]{mon88} Montgomery, D., \&
Phillips, L. 1988, \pra, 38, 2953
\bibitem[Ortolani \& Schnack(1993)]{ort93} Ortolani, S., \& Schnack, D.D. 1993, Magnetohydrodynamics of
Plasma Relaxation, World Scientific Publishing, Singapore.
\bibitem[Otto et al.(2007)]{ott07} Otto, A., B\"uchner, J., \&
Nikutowski, B. 2007, A\&A, 468, 313
\bibitem[Prigogine(1947)]{pri47} Prigogine, I.  1947,
{\textit{Etude Thermodynamique des Ph\'{e}nom\`{e}nes
Irr\`{e}versibles}}, Edition Desoer, Li\`{e}ge.
\bibitem[Schrijver et al.(2006)]{sch06} Schrijver, C.J.  et al.
2006, {{Sol. Phys.}}, {{235}}, 161
\bibitem[Shaikh et al.(2007)]{shk07} Shaikh, D., Dasgupta, B.,
Zank, G., \& Hu, Q. 2007, Phys. Plasmas, submitted
\bibitem[Song et al.(2006)]{son06} Song, M.T., Fang, C., Tang, Y.,
Wu, S.T., \& Zhang, Y. 2006, \apj, 649, 1084
    \bibitem[Taylor(1974)]{taylor74} Taylor, J. B. 1974,  \prl,
33, 1139
\bibitem[Venkatakrishnan \& Gary(1989)]{ven89} Venkatakrishnan, P., \&
    Gary, G. A.  1989, { Sol. Phys.}, 120, 235




\end{thebibliography}
\end{document}